\documentclass{ws-ijmpb}

\begin{document}

\markboth{A. Wagner}
{Spurious velocities in lattice Boltzmann}

\catchline{}{}{}

\title{The origin of spurious velocities in lattice Boltzmann}

\author{\footnotesize Alexander J. Wagner\footnote{
now at the Department of Physics, NDSU, Fargo, ND 58105-5566, USA}}

\address{Department of Physics \& Astronomy, University of Edinburgh,
JCMB King's Buildings, Mayfield Rd., Edinburgh EH9 3JZ, U.K.}

\maketitle

\pub{Received (August 5, 2002)}{Revised (revised date)}

\begin{abstract}
Stationary droplets simulated by multi-phase lattice Boltzmann methods
lead to spurious velocities around them.  In this article I report the
origin of these spurious velocities for one example and show how
they can be avoided.
\end{abstract}

\section{Introduction}
Imagine a stationary droplet in quiescent conditions. For a real
system there are, of course, no velocities. Yet when this simple
system is simulated with a lattice Boltzmann method
you will find a flow-field around the drop. Even if you start with a
no flow initial condition this velocity field will develop. An example
of these velocities in shown in Figure \ref{fig1} (a). These
velocities are know as ``spurious velocities''.

The magnitude of the velocities depends on the details of the method,
the radius of the drop, the surface tension and the viscosity. There
have been several studies\cite{Nourgaliev2002,Teng2000,Hou1997} of
different methods that tell us about the dependence of the spurious
velocities on these parameters. In the past the focus of the work has
been on reducing the magnitude of these velocities but to the best of
my knowledge no good understanding of their origin has been reached. In
this article I will explain why we see these spurious velocities at
all and how, and at what cost, they can be avoided.

I will first introduce a simple lattice Boltzmann method for model B
dynamics and show the perfect approach to equilibrium which is free of
any spurious currents. Then I extend this model to include
hydrodynamics and we will see that spurious currents suddenly appear.
A simple examination shows why there should not be any spurious
currents according to the continuum equations and I point out the
terms through which discretization errors drive the spurious currents.
I introduced a thermodynamically consistent discretization of these
terms which leads to a lattice Boltzmann method that is free of spurious
currents.

\section{Lattice Boltzmann for Model B}
Model B is a model describing the behavior of binary alloys. We will
consider a mixture of, say, $A$ and $B$ atoms. The system is
characterized by the order parameter $\phi$ which represents the
difference in the densities of the two components, {\it i.e.}
$\phi=-1$ represents pure $A$ and $\phi=1$ represents pure $B$.
The dynamics of this conserved order parameter $\phi$ is then given by
\begin{equation}
\partial_t \phi = \nabla D\nabla\mu; \;\;\;\;\;\;\;
\mu = \frac{\delta F}{\delta \phi}
\label{order}
\end{equation}
where D is a diffusion coefficient and $\mu$ is the chemical
  potential.
The system is described by the free energy $F$. For
simplicity we will consider a Landau Free energy expansion around the
critical point which is given by
\begin{equation}
F=\int \frac{A}{2}\phi^2+\frac{B}{4}\phi^4+\frac{\kappa}{2}(\nabla\phi)^2
\end{equation}
To simulate this equation we use a BGK lattice Boltzmann
method given by
\begin{equation}
g_i(\mathbf{x+v}_i,t+1) = g_i(\mathbf{x},t) 
	+ \frac{1}{\tau} [g^0_i(\mathbf{x},t) - g_i(\mathbf{x},t)]
\label{LBg}
\end{equation}
where the relaxation time $\tau$ may depend on $\phi$ and the set of
discrete velocities $\{\mathbf{v}_i\}$ will be determined later.  The
behavior of the model will be determined by the moments of the
equilibrium distribution $g^0_i$. We name these moments
\begin{equation}
\phi = \sum_i g_i^0 = \sum_i g_i;\;\;\;\;
\mathbf{F}   = \sum_i g_i^0 \mathbf{v}_i;\;\;\;\;
M = \sum_i g_i^0 \mathbf{v}_i \mathbf{v}_i.
\end{equation}
A Taylor expansion to second order in the derivatives
gives
\begin{equation}
(\partial_t +\mathbf{v}_i\nabla) g_i^0 
- (\partial_t +\mathbf{v}_i\nabla)
(\tau -\frac{1}{2})(\partial_t +\mathbf{v}_i\nabla) g_i^0
=  \frac{1}{\tau} [g^0_i - g_i]
\end{equation}
We now take the first moment ($\sum_i$) of this expansion and obtain
as the equation of motion for the order parameter $\phi$
\begin{equation}
\partial_t \phi + \nabla \mathbf{F} =  \nabla
(\tau-0.5) \partial_t \mathbf{F} + \nabla (\tau-0.5)\nabla M
\end{equation}
We have two independent ways to ensure that this equation be
equivalent to (\ref{order}). Either we choose $\mathbf{F} = 0,$ $M =
\mu\mathbf{1}$, $\tau=D+0.5$ or alternatively $\mathbf{F} = D\nabla \mu$
and $M = 0.$ Note that the $\nabla \partial_t\mathbf{F}$ term is now
third order. In two dimensions we can implement both of these
approaches with a small velocity set of five velocities
$\{\mathbf{v}_i\}=\{(0,0),(1,0),(-1,0),(0,1),(0,-1)\}$.

Both approaches behave equivalently, as expected, but they differ in
the ranges of stability as will be discussed elsewhere\cite{improve}.
Both implementations of this simple diffusive lattice Boltzmann
algorithm are free from spurious currents. We conclude that
hydrodynamics is required to see spurious currents.

\section{Model H -- binary fluids}
In order to simulate binary fluids we have to introduce a fluid
velocity which will convect the order parameter $\phi$. This fluid
velocity will obey the Navier-Stokes equation which is coupled to the
order parameter $\phi$ through a thermodynamics pressure tensor
$P^c$. We will denote the fluid density as $\rho$. The equations of
motion are\cite{improve}
\begin{eqnarray}
\rho \partial_t \phi + \rho \nabla (\phi \mathbf{u}) &=& \nabla D
\rho(1-\phi^2)\nabla \mu + \nabla D(\mu-\phi T)\nabla \rho
\label{orderc}\\
\partial_t \rho + \nabla (\rho \mathbf{u})&=&0\\
\rho \partial_t \mathbf{u} + \rho \mathbf{u} \nabla \mathbf{u}
&=& \nabla p + \nabla P^c 
+ \eta \nabla [\nabla \mathbf{u}+(\nabla
\mathbf{u})^T-\mbox{tr}(\nabla \mathbf{u})/d]
\end{eqnarray}
where the pressure is $p=\rho T$, $\eta$ is the viscosity and $d$ the
number of spatial dimensions.
The pressure tensor $P^c$ is defined as
\begin{equation}
P^c = [\phi \partial_\phi F - F 
-\kappa (\phi\nabla^2\phi+0.5\nabla\phi\nabla\phi)]\mathbf{1}
+\kappa \nabla \phi\nabla\phi
\end{equation}
We can implement these continuum equations using lattice
Boltzmann. With (\ref{LBg}) we now define $\sum_i g_i = \phi\rho$. By
imposing $\sum_i g_i^0 \mathbf{v}_i=\phi\rho \mathbf{u}+F$ and $\sum_i
g_i^0 \mathbf{v}_i \mathbf{v}_i = M + \phi\rho \mathbf{u}
\mathbf{u}$ we obtain the required order parameter equation
(\ref{orderc}) with either of the previous choices for $\mathbf{F}$
and $M$.
The Navier-Stokes equations are obtained with a usual one-component lattice
Boltzmann method which is coupled to the order parameter equation
through a body force of $\nabla P^c$.

\begin{figure}[htbp] 
\begin{minipage}{0.5\textwidth}
\begin{center}
\resizebox{0.9\textwidth}{!}{\includegraphics{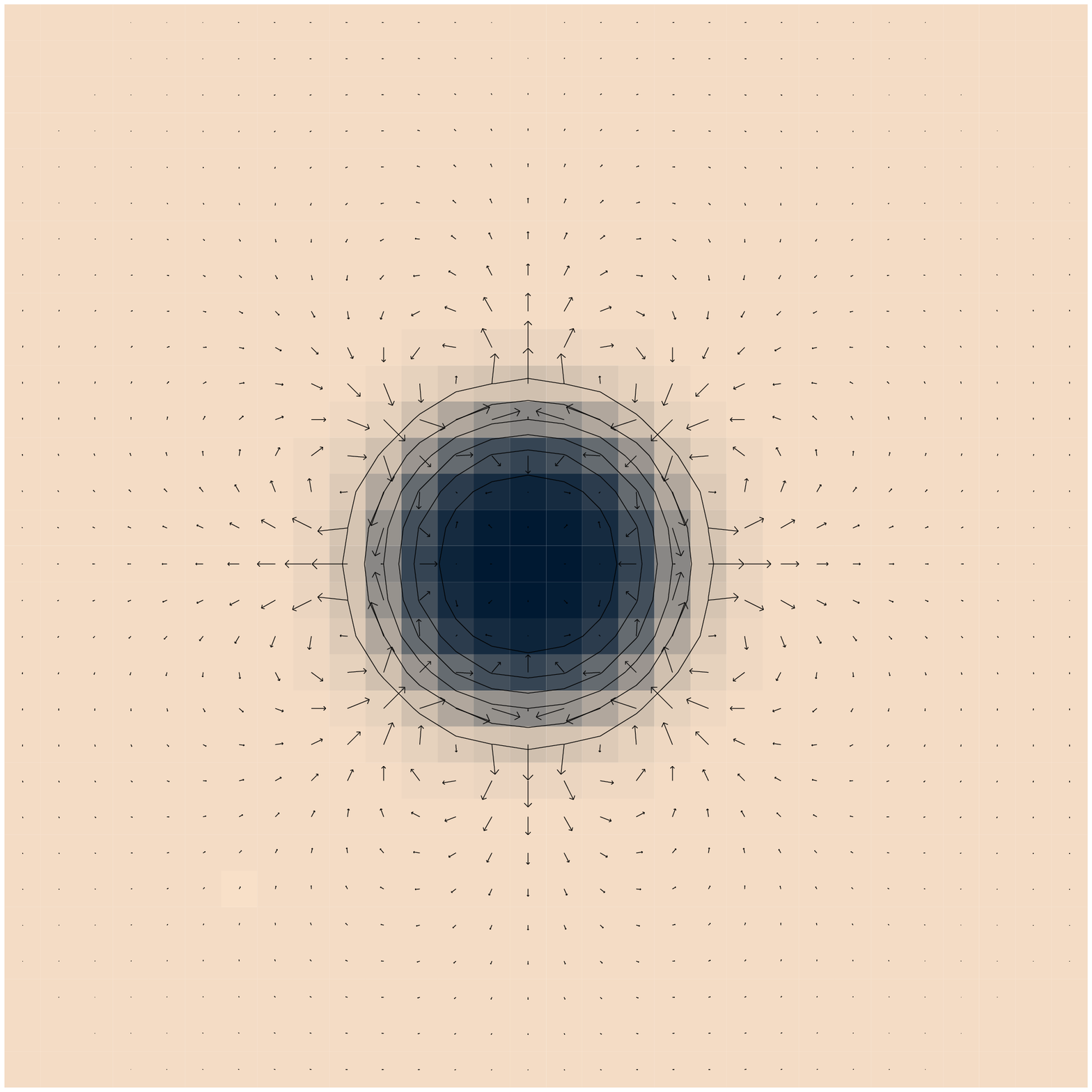}}\\
(a)\end{center}
\end{minipage}
\begin{minipage}{0.5\textwidth}
\begin{center}
\resizebox{0.9\textwidth}{!}{\includegraphics{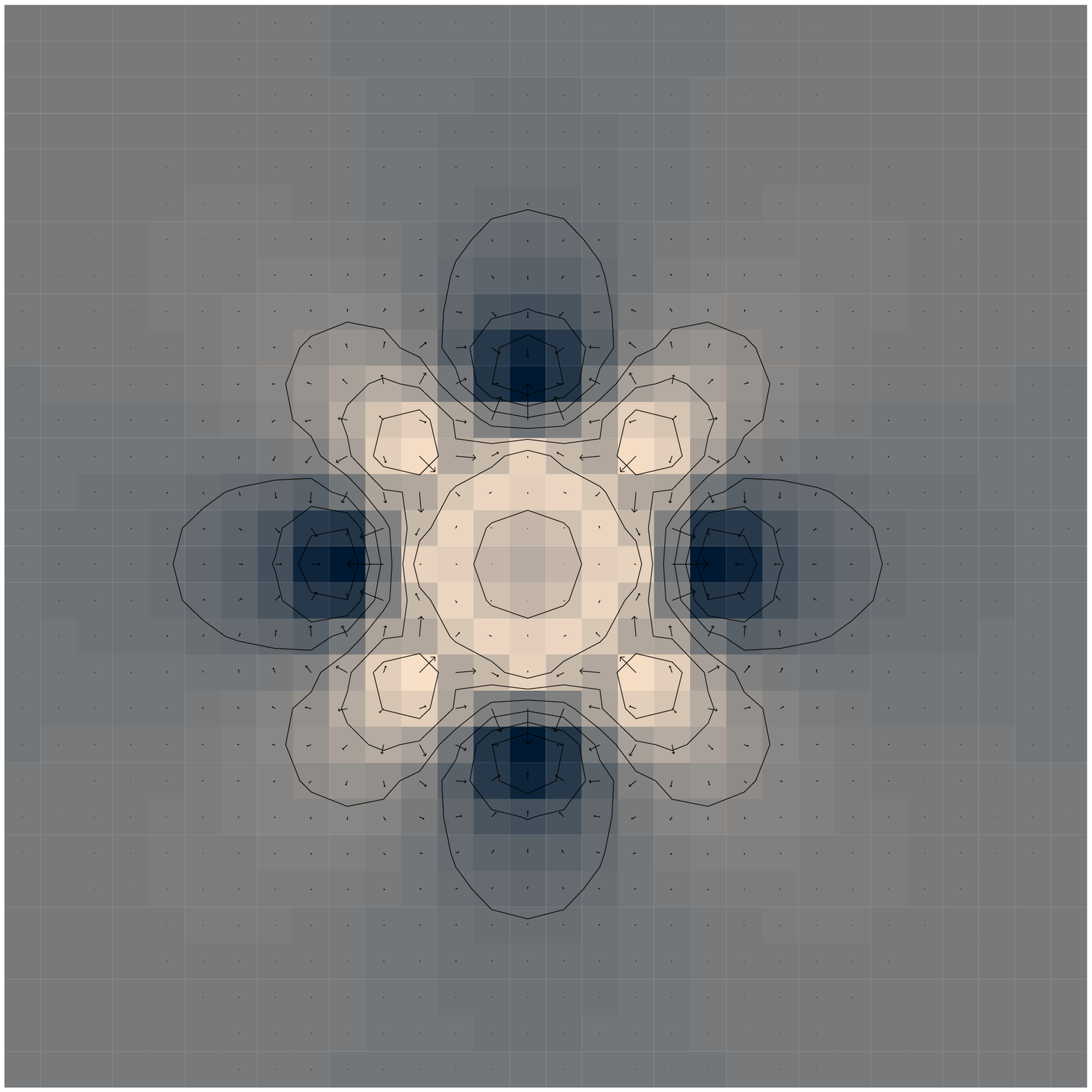}}\\
(b)\end{center}
\end{minipage}
\caption{(a) Spurious currents for a drop where the longest velocity
vector corresponds to a lattice velocity of $3.4\;10^{-4}$.The density
of the matrix is $\phi=0.931$ and the density of the drop is
$\phi=-1.013$. (b) The chemical potential and the spurious diffusion
currents for the same drop. The longest velocity vector corresponds to
a lattice velocity of $2.4\;10^{-4}$. The values of $\mu$ vary between
$-1.298\;10^{-3}$ and $-1.181\;10^{-3}$.}
\label{fig1}
\end{figure}

Doing this leads to the well known spurious velocities as shown in
Figure \ref{fig1}. We see that there are not only spurious currents
but also the chemical potential $\mu$ is no longer constant. Since the
implementation of a simple model B did not lead to the spurious
velocities it is reasonable to assume that the implementation of the
additional lattice Boltzmann equation for the total density and the
momentum is the reason for the occurrence of the spurious
velocities. Why should the continuous equations in
equilibrium be consistent with a quiescent drop without spurious
velocities? This is because the two driving terms $\nabla D\nabla \mu$
and $\nabla P^c$ are related through
\begin{equation}
\nabla \mathbf{P}^c = \phi \nabla \mu
\end{equation}
This means that a constant chemical potential will lead to zero
driving force in both equations.


When we now use this knowledge and replace the driving force $\nabla
P^c$ with $\phi \nabla \mu$ we find that the spurious velocities
vanish to machine precision. The density of the order-parameter in the
surrounding fluid is $\phi=0.933$ and the density in the drop is
$\phi=-1.015$ (cf Figure \ref{fig1}). The size of the maximum velocity
is now $5.6\;10^{-16}$. The chemical potential has the value of
$\mu=-1.2103\;10^{-3}$ and variations are smaller than
$2\;10^{-15}$. So why do the spurious velocities appear in the first
place? This is because the discretizations of $\nabla P^c$ and $\phi
\nabla \mu$ are very different. We conclude that the different
discretizations of the driving forces for the order parameter and
momentum equations are the origin of spurious velocities for lattice
Boltzmann.

There is, however, a caveat. The term $\phi \nabla \mu$ in its
discrete form is not a divergence of a scalar field which means that
momentum is now only approximately conserved. To be able to see the
absence of spurious velocities I included a tiny correction term in
the definition of the momentum that ensures that the total momentum of
the system does not change. Also my implementation of the algorithm
turned out to be unstable so I added a small amount of numerical
viscosity by multiplying a velocity by $(1-0.001)$ and adding 0.00025
times the velocity of the four nearest neighbors. This rendered the
simulations stable.

\section{Conclusions}
I proved that spurious velocities in one particular lattice
Boltzmann implementation are caused by non-compatible discretizations
of the driving forces for the order-parameter and momentum
equations. The different discretization errors for the Forces compete
and drive the spurious currents. I believe that the same argument holds
for all lattice Boltzmann methods that exhibit spurious
velocities. These spurious velocities can be avoided by ensuring that
the discretizations of the driving forces are compatible.\\
{\bf Acknowledgement} This work was funded by EPSRC GR/M56234 and GR/R67699.


\begin{thebibliography}{1}

\bibitem{Nourgaliev2002}
R. Nourgaliev, T. Dinh, and B. Sehgal, Nucl. Eng. Des. {\bf 211},  153  (2002).

\bibitem{Teng2000}
S. Teng, Y. Chen, and H. Ohashi, Int. J. Heat Fluid Flow {\bf 21},  112
  (2000).

\bibitem{Hou1997}
S. Hou {\it et~al.}, J. Comput. Phys. {\bf 138},  695  (1997).

\bibitem{improve}
A. Wagner, improvements for multiphase lattice Boltzmann methods (unpublished).

\end{thebibliography}
\end{document}